 \documentclass[authoryear,preprint,1p,12pt]{elsarticle}
 
 \makeatletter
\def\ps@pprintTitle{%
  \let\@oddhead\@empty
  \let\@evenhead\@empty
  \def\@oddfoot{\reset@font\hfil\thepage\hfil}
  \let\@evenfoot\@oddfoot
}
\makeatother

\usepackage{amssymb}
\usepackage{amsmath}
\usepackage{latexsym}
 \usepackage{graphicx}
\usepackage{setspace}

 \usepackage{booktabs}
 \usepackage{multirow}

\biboptions{sort}

\journal{}

\begin{document}

\begin{frontmatter}



\title{Composite random search strategies based on non-directional sensory cues }


\author[case]{Ben C. Nolting\corref{cor1}\fnref{label1}}
\ead{bcn13@case.edu}
\author[cram]{Travis M. Hinkelman\fnref{label1}}
\ead{travis.hinkelman@fishsciences.net}
\author[sbs]{Chad E. Brassil}
\ead{cbrassil@unl.edu}
\author[sbs]{Brigitte Tenhumberg}
\ead{btenhumberg2@unl.edu}

\address[case]{Department of Biology, 308 Clapp Hall, Case Western Reserve University, Cleveland, Ohio 44106-7080, USA}
\address[cram]{Cramer Fish Sciences, 13300 New Airport Road, Suite 102, Auburn, CA 95602, USA}
\address[sbs]{School of Biological Sciences, 348 Manter Hall, University of Nebraska, Lincoln, Nebraska 68588-0188, USA}
\cortext[cor1]{Corresponding author. Tel.: +1 402 613 1733; Fax: +1 216 368 467.}
\fntext[label1]{Contributed equally to this work}

\begin{abstract}
Many foraging animals find food using composite random search strategies, which consist of intensive and extensive search modes. Models of composite search can generate predictions about how optimal foragers should behave in each search mode, and how they should determine when to switch between search modes. Most of these models assume that foragers use resource encounters to decide when to switch between search modes. Empirical observations indicate that a variety of organisms use non-directional sensory cues to identify areas that warrant intensive search. These cues are not precise enough to allow a forager to directly orient itself to a resource, but can be used as a criterion to determine the appropriate search mode. As a potential example, a forager might use olfactory information as a non-directional cue. Even if scent is too imprecise for the forager to immediately locate a specific food item, it might inform the forager that the area is worth searching carefully. We developed a model of composite search based on non-directional sensory cues. With simulations, we compared the search efficiencies of composite foragers that use resource encounters as their mode-switching criterion with those that use non-directional sensory cues. We identified optimal search patterns and mode-switching criteria on a variety of resource distributions, characterized by different levels of resource aggregation and density. On all resource distributions, foraging strategies based on the non-directional sensory criterion were more efficient than those based on the resource encounter criterion. Strategies based on the non-directional sensory criterion were also more robust to changes in resource distribution. Our results suggest that current assumptions about the role of resource encounters in models of optimal composite search should be re-examined. The search strategies predicted by our model can help bridge the gap between random search theory and traditional patch-use foraging theory.
\end{abstract}

\begin{keyword} area-restricted search \sep composite search \sep giving-up time \sep L\'evy walk \sep ballistic motion \sep Brownian motion \sep optimal foraging



\end{keyword}

\end{frontmatter}

%%\linenumbers

\section{Introduction}
\label{intro}

For many organisms, the ability to efficiently find food resources is a key determinant of fitness \citep{Bell:1991uu}. It is advantageous for foraging animals to focus search effort on resource rich areas and minimize energy spent searching resource poor areas \citep{Viswanathan:2011vl}. This search tactic has been termed composite search \citep{Plank:2008gr}, area-restricted search \citep{Weimerskirch:2007it}, or area-concentrated search \citep{Benhamou:1992va}. A forager using a composite search strategy alternates between intensive and extensive search modes. In intensive mode, a forager thoroughly searches resource rich areas by making short moves and reorienting frequently; in extensive mode, it moves directly across resource poor areas by making long, relatively straight moves with minimal turning.

Composite search behavior is widespread, observed in taxa as diverse as slime moulds \citep{Latty:2009gc}, beetles \citep{Ferran:1994to}, honeybees \citep{Tyson:2011uz}, fish \citep{Hill:2003tf}, birds \citep{Nolet:2002wa}, ungulates \citep{Tyson:2011uz}, turtles \citep{Tyson:2011uz}, weasels \citep{Haskell:1997tx}, and humans \citep{hills2013ae}. Given the ubiquity of composite search, an important question arises: how should a forager determine when to switch from intensive to extensive mode, and vice versa? Questions about optimal foraging have traditionally been addressed with patch models that envision intensive search taking place within patches and extensive search as movement between patches \citep{Charnov:1976wz,Oaten:1977vs}. These models are not directly applicable to cases where resources do not occur in well-defined patches, and instead take on more general spatial distributions \citep{Arditi:1988vf}. Optimal foraging on such landscapes is more properly addressed using random search theory \citep{Viswanathan:2011vl,James:2010ic,Reynolds:2009ub}. In random search models, resources are represented as points, and animal movement is modeled with stochastic processes. Unlike patch models, random search models are spatially explicit; resource locations in these models can be specified according to any spatial point pattern and are not limited to the case of clearly defined patches.

Recently, many studies have compared the efficiencies of different random search movement patterns \citep{James:2008bs,James:2011kw,Reynolds:2009gs}, and composite searches have been a particular focus \citep{Reynolds:2010jn,Plank:2008gr,Reynolds:2009es, Benhamou:2007vka}. The criteria that foragers use to switch between modes have received far less attention. Most analyses of optimal composite search presume that foragers use a ``giving-up time'' (GUT) as their mode-switching criterion \citep{Reynolds:2010jn,Plank:2008gr,Reynolds:2009es,Scharf:2007gf}. A forager using this criterion switches from extensive to intensive mode upon encountering a resource. It then stays in intensive mode until a fixed amount of time (the GUT) has elapsed without a subsequent resource encounter. GUT models accurately describe some foraging situations, such as ladybird beetle larvae (\textit{Coccinella septempunctata}) feeding on aphids \citep{Carter:1982vi} and houseflies (\textit{Musca domestica}) feeding on sucrose drops \citep{Bell:1990tg}.

Rather than keeping track of time, many animals use sensory cues to modulate their search behavior. Parasitoids like \textit{Nermeritis canecens} \citep{Waage:1979wx}, \textit{Venturia canescens} \citep{Bell:1990tg}, and \textit{Cardiochiles nigriceps} \citep{Strand:1982uz} use chemical cues to determine when to search intensively for hosts. When deciding when to leave a foraging site, wolf spiders rely more heavily on visual and vibratory cues than elapsed time since their last prey encounter \citep{Persons:1997td}. Procellariiform seabirds use chemicals like dimethyl sulfide to identify when to switch to intensive search behavior (in this case, intensive search consists of upwind zig-zag movement) \citep{nevitt2008pn}. These seabirds ``use changes in the olfactory landscape to recognize potentially productive foraging opportunities as they fly over them" \citep{nevitt2008sy}. Further examples of animals that use sensory cues to determine search mode include ciliates like \textit{Paramecium} and \textit{Tetrahymena} \citep{Levandowsky:1988vkb,Leick:1992ty}, bacteria, like \textit{Escherichia coli} and \textit{Salmonella typhimurium} \citep{Adler:1975wi,Moore:2004ky,Dusenbery:1998tq}, cod larvae \citep{Doving:1994tn}, and fruit flies \citep{DalbyBall:2000uj}. It is important to note that identifying discrete behavioral states (e.g., search modes) from empirical movement data is a difficult problem; fortunately, significant progress has been made in this area \citep{Nams:2014cu, Postlethwaite:2013jw, Knell:2011km, Barraquand:2008tc}.

There are two primary ways that organisms use sensory cues to find resources: taxis and kinesis \citep{Codling:2008rw,dusenbery1989ef}. In taxis, an organism uses sensory cues (e.g., the gradient of a stimulus field) to orient itself and move toward the resource. In kinesis, an organism uses sensory cues to determine its speed (orthokinesis) or turning frequency (klinokinesis). Unlike taxis, kinesis does not use directional information. Taxis is an efficient strategy used by many organisms \citep{Seymour:2010ca}, but in some situations limitations on sensory information make it impractical; \citet{Hein:2012bd} note that such limitations occur when sensory signals are infrequent, noisy, or contain limited directional information. When organisms are unable to extract directional information from sensory cues, they may still be able to use kinesis. In this paper, we refer to the cues used in kinesis as \textit{non-directional sensory cues}. We use this term to contrast with \textit{directional sensory cues}, which allow foragers to orient their motion toward resources. Most foragers likely use a combination of non-directional and directional sensory cues. For example, a forager might use odor as a non-directional cue to determine where to search intensively; when it comes close to a resource, it might switch to using visual information as a directional cue and move directly to the resource. A forager that uses odor as a non-directional cue when the signal is dilute and the odor gradient is imperceptible might switch to taxis (using odor as a directional cue) when it is close to a resource and the odor gradient is more pronounced. Two specific examples illustrate how foragers use non-directional sensory cues. \citet{williams1994ms} proposes that tsetse flies search for targets using kinesis, with carbon dioxide concentration serving as a non-directional sensory cue. Williams hypothesizes that winds in typical tsetse fly habitats are too light and variable to allow for taxis based on wind direction. Juvenile flatfish use kinesis to find bivalves \citep{hill2000id, hill2002ae}; respiratory currents generated by the bivalves are likely the non-directional sensory cue in this system.

In this study, we model two classes of composite search strategies: those with mode transitions triggered by resource encounters and elapsed time (the GUT criterion), and those with mode transitions triggered by non-directional sensory cues. Our modeling framework allows for a full spectrum of random movement patterns for both intensive and extensive mode. We used simulations to compare the efficiencies of different search strategies. Searching efficiency depends in part on the spatial distribution of resources \citep{Cianelli:2009ti}, so we compared search strategies on a variety of landscape types, characterized by different levels of resource aggregation and density. Further, we examined the performance of the search strategies in response to changes in resource aggregation to test the robustness of the search strategies to environmental change. We found that the search strategy based on non-directional sensory cues outperformed the search strategy based on resource encounters across all landscape types, and was more robust to changes in resource aggregation.

\section{Methods}
\label{sec:1}
\subsection{Overview}

In our modeling framework, resources are represented as points distributed across a two-dimensional landscape, and a forager is represented as a moving point with a small fixed detection radius. The forager moves at a constant speed, and when a resource falls within its detection radius, the forager moves in a straight line to the resource and consumes it; otherwise, the forager implements a random search strategy. Random search strategies consist of a set of probabilistic movement rules. Although the resulting movement patterns are stochastic, the probability distributions that generate the movement provide a structure for the search.

In accordance with many theoretical studies on optimal random search behavior (e.g., \citealt{Viswanathan:1999vc, Reynolds:2010jn, James:2008bs}), our model is very general, and parameters are not fit to any particular species. The distance and time units in our simulation set the characteristic distance and time scales of the system. These units could be quantified in terms of meters and seconds to represent a specific system. Our simulations use a square landscape 101 units in length and width, and foragers have a detection radius of 0.5 units. Coordinates are floating point numbers, and are not restricted to discrete values. Like many random search simulations (e.g., \citealt{Reynolds:2009es}), we focus on a case where the detection radius is less than 0.01 of the landscape scale.

\subsection{L\'evy walks}

L\'evy walks are stochastic processes that provide a versatile tool for modeling animal movement \citep{Bartumeus:2013vm, Reynolds:2009ub, Viswanathan:2011vl}. A L\'evy walk with parameter \(\mu\) is a random walk with step lengths \(l\) drawn from a Pareto distribution, \(p(l)\sim l^{-\mu}\), \(1<\mu \leq 3\). Different values of \(\mu\) produce different types of random walks. As \(\mu \rightarrow 1 \), the resulting random walk approaches ballistic (i.e., straight-line) motion. Random walks with step lengths drawn from a Pareto distribution with \(\mu \geq 3\) behave like Brownian motion. Thus, L\'evy walks can be seen as spanning a spectrum of movement behavior, ranging from ballistic motion (\(\mu \rightarrow 1 \)) on one extreme to Brownian-like motion (\(\mu=3\)) on the other (for details see, \ref{sec:levyjust}). This family of random walks is a widely used modeling tool (e.g.,\citealt{Viswanathan:1999vc}).

L\'evy walks are not to be confused with L\'evy flights. In the former, a forager moves continuously along each step length; in the latter, a forager hops instantly from the start to the end of each step length. L\'evy walks model cruise foragers, while L\'evy flights model saltatory foragers (for more on this distinction, see \citealt{James:2010ic}).

Most L\'evy walk models, including those considered in this study, are technically truncated L\'evy walks: step lengths are terminated when a resource or boundary is reached, or when the maximum time of the simulation elapses \citep{Reynolds:2009ub}. Fortunately, many of the important features of L\'evy walks, including general properties of the mean-square displacement, are retained by truncated L\'evy walks \citep{Viswanathan:2008in, Mantegna:1994vc}. For more details, see \ref{sec:levyjust}.

Our model deals with both non-composite and composite foragers. Non-composite foragers move by L\'evy walks with parameter \(\mu\). Composite foragers switch between extensive and intensive search modes. In extensive search mode, foragers move according to a L\'evy  walk with parameter \(\mu_\text{ext}\). In intensive search mode, foragers move according to a L\'evy walk with parameter \(\mu_\text{int}\).  Previously, composite searches have been modeled with Brownian motion in the intensive mode and ballistic motion in the extensive mode \citep{Plank:2008gr, Benhamou:2007vka}. This was later generalized to consider a full range of L\'evy walks in extensive mode \citep{Reynolds:2009es}. Our model represents a further generalization that allows a full range of L\'evy walks for both intensive and extensive search modes.

Correlated random walks provide another approach to modeling movement on the ballistic to Brownian spectrum. This approach has been used with great success \citep{Benhamou:2007vka, Benhamou:2013de}. The L\'evy walk and correlated random walk approaches are compatible, and are mathematically connected \citep{Reynolds:2010cl}. We used L\'evy walks to build our models, because they offered a straightforward way of implementing non-directional sensory mode-switching criteria (as described in section 2.4). Future studies should examine how our models can be translated into a correlated random walk framework.

A recent review by \citet{Pyke:2014gx} suggests that the enthusiasm with which many ecologists have embraced L\'evy walks is misplaced. In the discussion, we address how his criticisms of the L\'evy walk concept apply to our model.

\subsection{Forager movement}

In our model, a forager moves by selecting a heading and a step length. The heading is randomly selected from a uniform distribution on $\left[0,\ 2\,\pi\right)$. The step length is selected from a Pareto distribution with parameter \(\mu\) (for a non-composite forager), \(\mu_\text{int}\) (for a composite forager in intensive mode), or \(\mu_\text{ext}\) (for a composite forager in extensive mode). The procedure for simulating ballistic motion was an exception that will be described at the end of this subsection.

For non-ballistic motion, the selected heading and step length together determine a random walk step. The forager moves along a random walk step at a uniform speed of 0.25 units per time step. The forager's speed determines how finely its movement is discretized, and 0.25 was the smallest speed that allowed for practical simulation. It takes a forager many time steps to complete a typical random walk step.

If the forager encounters a resource while it is moving along a random walk step, it truncates the random walk step, moves to the resource, and consumes the resource. Consumed resources are not replaced; hence our simulations represent destructive foraging (resource depletion). If a forager reaches a landscape boundary before completing a random walk step, it truncates the random walk step (details on boundary conditions are provided in section 2.5). When a forager ends a random walk step, whether that step is truncated or not, it randomly selects another heading and step length, and the procedure repeats.

Simulations of ballistic motion (\(\mu \rightarrow 1 \)) do not use Pareto distributions to generate step lengths. A forager using ballistic motion selects a heading and moves in that direction until it encounters a resource or landscape boundary.

\subsection{Mode-switching criteria}

Our model considers two type of composite foragers: GUT foragers, which use resource encounters as their search mode criterion, and sensory foragers, which use non-directional sensory cues as their search mode criterion. A GUT forager switches from extensive to intensive search when it encounters a resource. After encountering a resource, the forager reverts to extensive search as soon as a specified time (the GUT) elapses without a subsequent resource encounter.

For sensory foraging, we created a generalized non-directional sensory field. We denote the intensity of non-directional sensory cues generated by a resource located at \(y_i\) detected at a location \(x\) by $f_{y_{i}}\left(x\right)$. The shape of the function $f_{y_{i}}\left(x\right)$ will depend on the particular sensory mechanisms involved; here, in order to make the model as general as possible, we assume that the strength of non-directional sensory cues generated by a resource at \(y_i\) follow the probability density function of a Gaussian distribution with variance \(\sigma^2\) centered at  \(y_i\). This is particularly appropriate if, for example, the sensory cues are chemical signals that travel via diffusion. For more specific cases, other distributions could be used to generate the non-directional sensory field (for example, the sensory field could be designed to account for how advection transports chemical cues or how vibratory signals are propagated through various media). The total non-directional sensory field is obtained by superimposing the fields produced by each resource, \(f\left(x\right)=\sum_{i}f_{y_{i}}\left(x\right)\). When a resource is consumed, its contribution to the non-directional sensory field is removed.

\begin{figure}
  \centering
  \includegraphics[width=5.12 in]{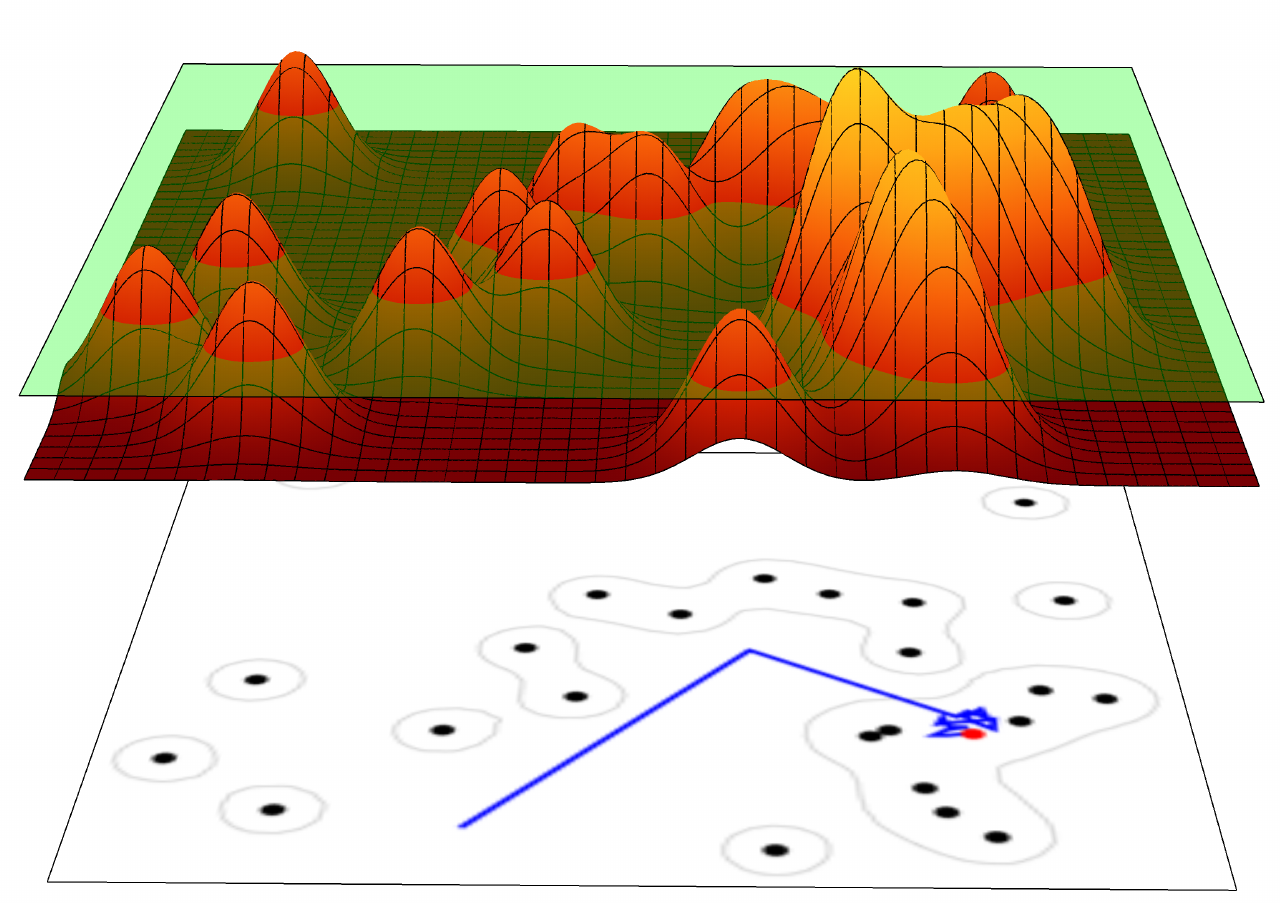}
  \caption{
  \bgroup
  \baselineskip=2pt
  A schematic representation of the behavior of a non-directional sensory forager. Resources are black dots on the two-dimensional landscape (bottom). The radius of a dot is the forager's detection radius. A non-directional sensory field (red surface) is generated by the resources. A non-directional sensory forager has a fixed threshold (green plane) that it uses as a mode-switching criterion. When a forager reaches the end of a step-length, it assess the sensory field; if the field is above the threshold value (circled areas on landscape), the forager engages in intensive search. The forager's movement is represented by the blue line. In this case, it eventually consumes a resource (red disk).
  \egroup }
\label{fig:sensoryfig}
\end{figure}

A non-directional sensory forager assesses the sensory field at the end of every random walk step (note that it does not assess the sensory field after every time step). If the value of the field is below a specified threshold, the forager engages in extensive search; if it is above the threshold, it engages in intensive search (Fig. \ref{fig:sensoryfig}). We call this threshold the sensory field threshold and abbreviate it as SFT. A forager has more opportunities to assess the sensory field in intensive mode (which is composed of many short random walk steps) than in extensive mode (which is composed of a few long random walk steps). This assumption reflects a trade-off between the attention devoted to movement versus monitoring the sensory field. 

It is important to note that the sensory forager and the GUT forager each have distinct advantages. The sensory forager has the benefit of using non-directional cues, but the GUT forager has the benefit of using elapsed time since its last resource encounter. It is not \textit{a priori} obvious which of these abilities leads to higher foraging efficiency. In reality, many foragers probably use a combination of sensory and GUT strategies. In this study, we focus on comparing these strategies in isolation; future work could examine how these strategies can be used together.

\subsection{Landscape characteristics}

Resources were distributed across landscapes according to Neyman-Scott processes \citep{Illian:2008uqa}. We selected this point process because it allowed us to adjust both the intensity and aggregation of the process. The resource distributions were specified by two parameters: the radius of the clusters of resources and the total initial number of resources. We used 100, 400, 700, and 1000 as our initial resource levels, and cluster radii of 4, 8, 16, 32, and 64.

The algorithm for generating realizations began with randomly drawing the number of resource aggregations, or clusters, from a Poisson distribution with an expected value of 15 (Table \ref{tab:parmvals}). The center of each cluster was randomly assigned to a point in the landscape (i.e., parent point). Then resources were sequentially assigned to a random parent and randomly placed within a specified radius (i.e., cluster radius) of the parent point until all resources were distributed among the parents. Thus, for each run of the simulation, the algorithm randomly determined the number of clusters and the number of resources per cluster, but the initial total resource density and the cluster radius were fixed. By changing a single parameter (i.e., cluster radius), we were able to vary the degree of aggregation of resources, which ranged from tightly clumped (cluster radius = 4) to dispersed (cluster radius= 64).

\begin{table}
 \caption{Parameter values used in the simulation model}
 \vspace{2 mm} 
  \centering
 {\footnotesize \begin{tabular}{l l}
  \toprule
    Parameter & Value
    \\ \midrule
    Resources & \\
    \(\hspace{5mm}\) Initial number of resources & 100, 400, 700, 1000 \\
    \(\hspace{5mm}\) Number of clusters\(^1\) & 15 \\
    \(\hspace{5mm}\) Radius of resource cluster\(^2\) & 4, 8, 16, 32, 64 \\
    Forager & \\
    \(\hspace{5mm}\) Speed (distance/time step) & 0.25 \\
    \(\hspace{5mm}\) Detection radius & 0.5 \\
    \(\hspace{5mm}\) L\'evy exponent (\(\mu\)) & \\
    \(\hspace{10mm}\) Extensive search mode & 1.0, 1.2, 1.4, 1.6, \(\ldots\), 3.0 \\
    \(\hspace{10mm}\) Intensive search mode & 1.0, 1.2, 1.4, 1.6, \(\ldots\), 3.0 \\
    \(\hspace{5mm}\) Mode-switching criteria\(^3\) & \\
    \(\hspace{10mm}\) Giving-up time & 0, 50, 100, 150, 200, \(\ldots\), 500 \\
    \(\hspace{10mm}\) Sensory field threshold & 0, 0.0005, 0.001, 0.002, \(\ldots\), 0.128, 0.256 \\
    \bottomrule
  \end{tabular}
  \label{tab:parmvals}
 \begin{quote}
\(^1\)Poisson random variable with an expected value of 15\\ 
\(^2\)Resource aggregation decreases with increasing cluster radius\\
\(^3\)Forager employs only one mode-switching criteria in a run of the simulation\\
\end{quote}}
\end{table}

A common misconception is that the negative binomial distribution is the best tool for modeling clusters. A negative binomial distribution describes the probability of finding a specific number of points within a sample area; it does not directly generate the positions of points. In fact, there is no stationary spatial point process that generates a negative binomial distribution of points in all possible sample areas \citep{Diggle:2003tz}. In contrast, the Neyman-Scott process is a stationary spatial point process.

The boundary conditions for the landscape were selected to minimize the impact of boundary artifacts. A buffer zone five units wide was added to each side of the landscape. This buffer zone contained no resources. Its purpose was to make sure that no resources were extremely close to the landscape boundary (which would protect them from approach from one or more sides). When a forager reached a boundary, it was relocated to a random position in the landscape, and it resumed its search (starting by drawing a new step length). The rationale for this type of boundary condition is described in \ref{sec:boundary}.

\subsection{Optimization}

We used Netlogo \citep{NetlogoCenterfor:1999ww} to simulate three classes of foraging strategies: non-composite, GUT, and non-directional sensory. Within each of these strategy classes, we sought to identify the movement parameters and mode-switching threshold that maximized search efficiency. We defined efficiency as the number of resources consumed divided by the length of the forager's trajectory.  For the non-composite foragers, this amounted to optimizing the movement parameter \(\mu\). For GUT foragers, we optimized \(\mu_\text{int}\), \(\mu_\text{ext}\), and the GUT. For non-directional sensory foragers, we optimized \(\mu_\text{int}\), \(\mu_\text{ext}\), and the SFT. Using an optimization algorithm (see \ref{sec:optimize}), we found the optimal parameter combination for each class of forager on each type of landscape, and compared the efficiencies of these optimal foragers. Then, we examined the sensitivity of search efficiency to each of the optimized parameters (see \ref{sec:sensitivity}). We also explored how a forager optimized to one type of landscape would fare in another; we quantified this ability with a measure called robustness (see \ref{sec:robustness}). The sensitivity and robustness analyses were conducted with R \citep{RAlanguageanden:2011vt}.

\section{Results}
\label{sec:3}

\subsection{Optimal parameters}

The optimal parameter for non-composite search generally ranged from \(\mu = 1.0\) (ballistic motion) on landscapes with low resource aggregation to \(\mu = 1.8\) on landscapes with high resource aggregation (Table \ref{tab:optparms}). Although optimizing the parameter for non-composite L\'evy walks is a well-studied problem, the case of destructive foraging on patchily distributed resources is not; such situations were once assumed to be equivalent to non-destructive foraging on uniform landscapes, but this is not true \citep{Reynolds:2010jn}. Our non-composite results are largely in agreement with previous results about destructive searches on landscapes generated by cellular automata \citep{Reynolds:2010jn}.

The optimal search parameters for composite foragers showed several interesting patterns. For all degrees of resource aggregation, the best GUT foraging strategies involved ballistic motion in extensive mode (\(\mu_\text{ext} = 1\)) (Table \ref{tab:optparms}). The optimal intensive mode for GUT foragers depended on the degree of resource aggregation. On landscapes with a high degree of resource aggregation, optimal GUT foragers used Brownian motion in intensive mode (\(\mu_\text{int} = 3\)). The optimal GUT foragers for other landscapes used a ballistic extensive strategy and a superdiffusive intensive strategy (\(\mu_\text{int} < 3\)). For all degrees of resource aggregation, the best non-directional sensory foraging strategies involved Brownian motion in intensive mode (\(\mu_\text{int} = 3\)). The optimal non-directional sensory foragers used an extensive mode that depended on the landscape, although these extensive modes were always ballistic or close to ballistic.

\begin{table}
 \caption{Parameter combinations for three different search strategies producing the highest mean searching efficiency for different resource densities and cluster radii. Resource aggregation decreases with increasing cluster radius.} 
  \vspace{2 mm} 
  \centering
  {\footnotesize \begin{tabular}{| c | c | c | c c c | c c c |}
  \hline
    Resource
    & \multicolumn{1}{c|}{Cluster}
    & \multicolumn{1}{c|}{NCS\(^1\)}
    & \multicolumn{3}{c|}{GUT Strategy}
    & \multicolumn{3}{c|}{NDS Strategy\(^2\)}
    \\ \cline{3-9}
    Density & Radius & \(\mu\)  & \(\mu_\text{ext}\)  & \(\mu_\text{int}\) & GUT & \(\mu_\text{ext}\)  & \(\mu_\text{int}\)  & SFT\(^3\) \\ 
    \hline
    100 & 4 & 1.6 & 1.0 &	3.0 & 250 &	1.2 & 3.0 &	0.0005 \\
    100 & 8 & 1.4	& 1.0 &	3.0	& 400 &1.4	& 3.0 &	0.0005 \\
    100 &16 &	1.2	&1.0&	2.6&	250	&1.6&	3.0	&0.0005 \\
    100 &	32 &	1.4	& 1.0&	1.8&	150	&1.4&	3.0&	0.0005 \\ 
    100 &	64 &	1.2 &	1.0	&1.4&	100	&1.6	&3.0&	0.0005 \\ 
    \hline
    400 &	4 &	1.6	&1.0	&3.0&	150	&1.2	&3.0	&0.0005 \\
    400&	8	&1.6&	1.0	&3.0	&150	&1.2	&3.0	&0.0020\\
    400	&16&	1.4&	1.0	&2.6	&150&	1.0	&3.0	&0.0010\\
   400&	32	&1.2&	1.0	&2.0	&100	&1.0	&3.0	&0.0010\\
   400&	64	&1.2&	1.0	&1.6	&50	&1.2	&3.0	&0.0040\\ 
   \hline
   700&	4	&1.6&	1.0	&3.0	&100	&1.2	&3.0	&0.0020\\
   700	&8	&1.4&	1.0	&3.0	&100	&1.0	&3.0	&0.0010\\
   700	&16&	1.4	&1.0&	2.6	&50	&1.2	&3.0	&0.0160\\
   700	&32&	1.2	&1.0&	2.0	&50	&1.0	&3.0	&0.0320\\
  700	&64&	1.0	&1.0&	1.0	&---	&1.0	&3.0	&0.0320\\ 
  \hline
1000	&4	&1.8	&1.0&	3.0	&100	&1.0	&3.0	&0.0005\\
1000	&8	&1.6	&1.0&	3.0	&100	&1.0	&3.0	&0.0005\\
1000	&16&	1.4	&1.0&	2.4	&50	&1.0	&3.0	&0.0320\\
1000	&32&	1.4	&1.0&	2.0	&50	&1.0	&3.0	&0.0640\\
1000	&64&	1.0	&1.0&	1.0	&---	&1.0	&2.8	&0.0640\\
    \hline
  \end{tabular}
  \label{tab:optparms}
  \begin{quote}
\(^1\)Non-composite search strategy\\
\(^2\)Non-directional sensory search strategy\\
\(^3\)Sensory field threshold
\end{quote}}
\end{table}

The optimal parameters identified in our simulations can be compared with conventional composite search strategies, which use ballistic motion in extensive search and Brownian motion in intensive search \citep{Plank:2008gr}. Optimal GUT foragers for landscapes with a high degree of resource aggregation behaved like a conventional composite searcher. The optimal GUT foragers for other landscapes used the conventional extensive strategy but deviated from the conventional intensive strategy (\(\mu_\text{int} < 3\)). The optimal non-directional sensory foragers used intensive and extensive movement parameters that are consistent with conventional composite search (although the criteria they use for mode-switching distinguishes them from previous composite search models).

\subsection{Search strategy comparisons}

\begin{figure}
  \centering
  \includegraphics[width=5.4 in]{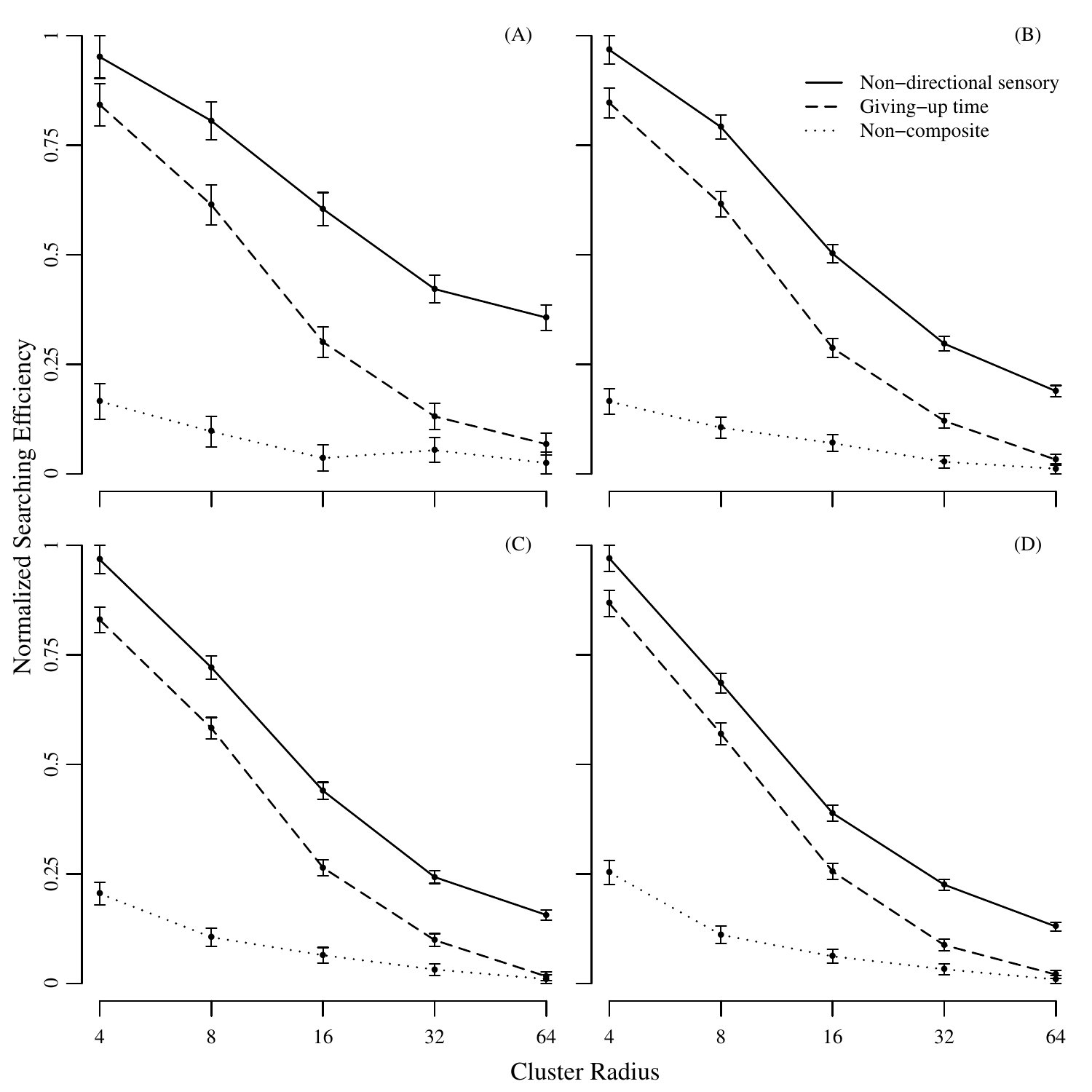}
  \caption[Normalized searching efficiency for three search strategies]{Normalized searching efficiency for three search strategies across 5 levels of resource aggregation (measured by cluster radii) and 4 levels of resource density: (A) 100, (B) 400, (C) 700, (D) 1000. Searching efficiency was normalized for comparison across resource densities. Error bars represent 95\% confidence intervals. Non-directional sensory search (solid lines) outperforms GUT search (dashed lines) across all landscape types. On landscapes with low aggregation (large cluster radii), the advantage of GUT over non-composite search (dotted lines) vanishes, but the advantage of non-directional sensory over non-composite search does not.}
\label{fig:normeff}
\end{figure}

\begin{figure}
  \centering
  \includegraphics[width=6 in]{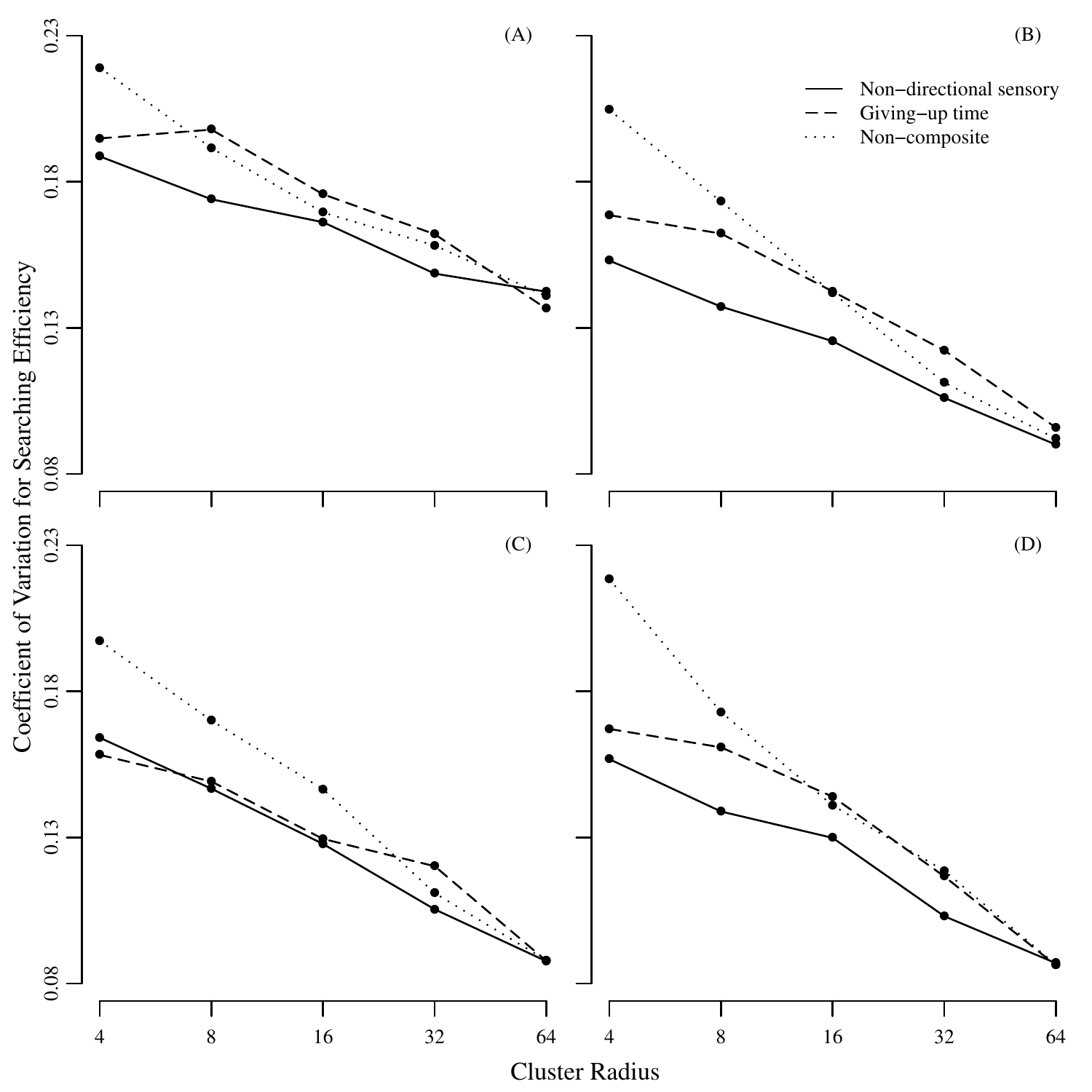}
  \caption{Coefficient of variation in searching efficiency for three search strategies across 5 levels of resource aggregation (measured by cluster radii) and 4 levels of resource density: (A) 100, (B) 400, (C) 700, (D) 1000. Composite search (solid and dashed lines) is less variable than non-composite search (dotted lines) on most landscape types, particularly on landscapes with high resource aggregation (low cluster radii). Non-directional sensory search (solid lines) is less variable than GUT search (dashed lines) on most landscape types.}
\label{fig:cv}
\end{figure}

After identifying optimal parameters for non-composite, GUT, and non-directional sensory foragers, we compared the search efficiencies of these foraging strategies. The composite search strategies outperformed the non-composite search strategy when resources were highly aggregated, and the relative advantage of composite search increased with the degree of resource aggregation (Fig. \ref{fig:normeff}). Composite search also produced lower variability in search efficiency than non-composite search when resources were aggregated (Fig. \ref{fig:cv}). For all search strategies, both search efficiency (Fig. \ref{fig:normeff}) and variability in search efficiency (Fig. \ref{fig:cv}) increased with degree of resource aggregation.

The non-directional sensory strategy performed better than the GUT strategy across the full spectrum of resource aggregation (Fig. \ref{fig:normeff}).  At first glance, this result may seem obvious; having sensory capabilities is clearly better than not having them at all. Recall, however, that the non-directional sensory forager is not simply an enhanced GUT forager. The GUT forager has the ability to keep track of time since the last resource encounter, an ability that the non-directional sensory forager lacks.

The non-directional sensory forager's performance advantage over the GUT forager can be attributed to two main reasons. First, the sensory forager has more opportunities to switch search mode. The GUT forager only switches mode upon encountering resources or when the time threshold expires. The sensory forager examines the sensory field at every resource encounter and at the end of every step of its random walk; this happens very frequently when move lengths are short (i.e., when \(\mu\) is close to 3.0). When the sensory forager engages in intensive mode, it is not making a large time commitment, because it has frequent opportunities to revert to extensive mode. When the GUT forager engages in intensive search, it is stuck in that mode until the time threshold elapses. Second, the GUT foragerÕs search strategy relies on the spatial autocorrelation of resources. When a GUT forager encounters a resource, it enters intensive search, under the assumption that other resources are nearby. In contrast, the sensory forager can be triggered into intensive search by local deviations in the sensory field, which is beneficial regardless of the spatial autocorrelation of the resources. This effect is evident in Figure \ref{fig:normeff}, where the advantage of sensory search over GUT search increases slightly as landscapes become more dispersed.

These results regarding the relative efficiencies of different search strategies have two major biological implications. First, they indicate that non-directional sensory strategies are generally superior to GUT strategies. From a forager's point of view, this means that it is better to keep track of sensory cues than time, even if the sensory cues cannot be used for taxis. Second, these results show that non-directional sensory foraging is a useful strategy, even when resources are not aggregated. This means that the advantages of composite search extend beyond environments with clumped resource distributions.

\subsection{Sensitivity}

\begin{figure}
  \centering
  \includegraphics[width=5 in]{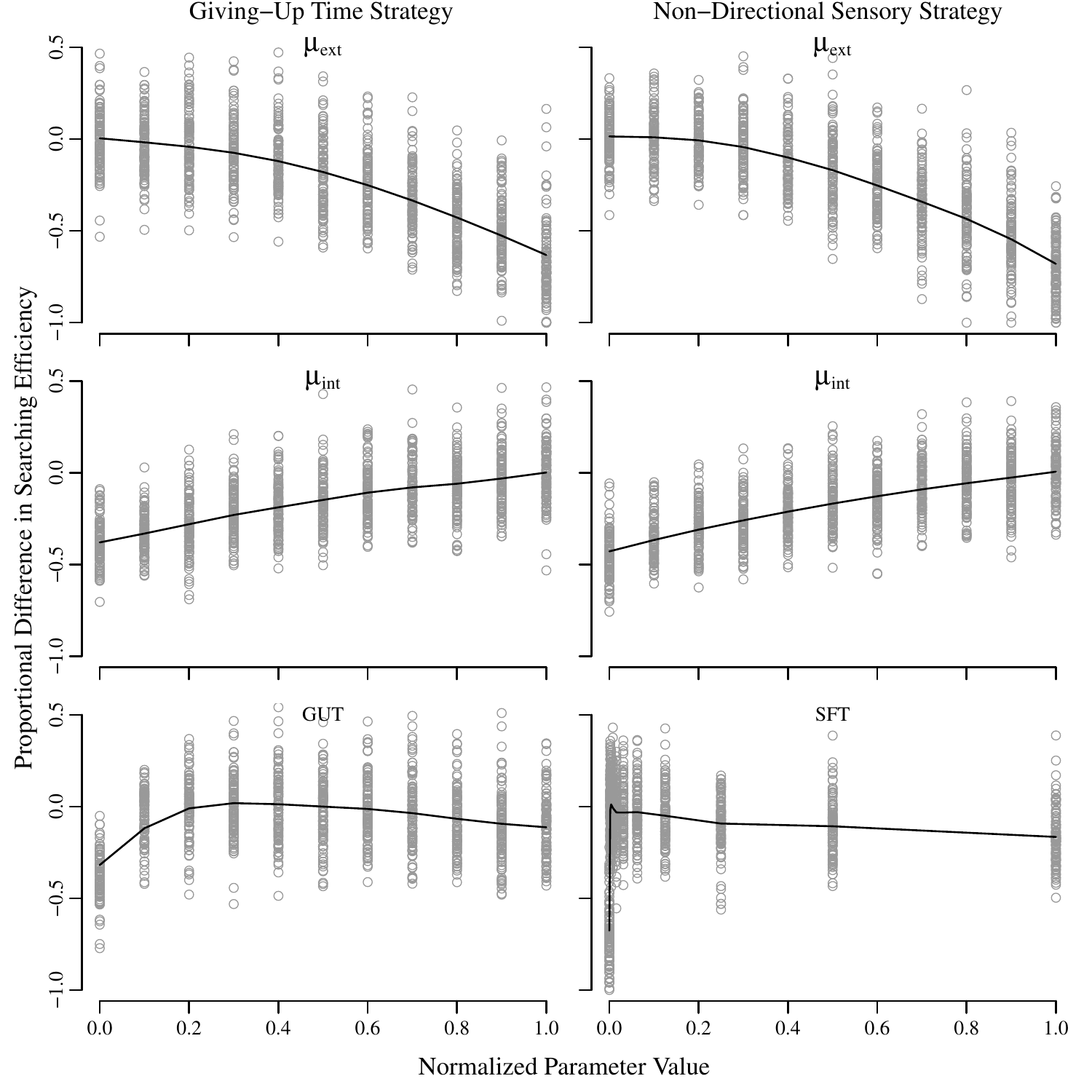}
  \caption{Representative example of sensitivity analysis for the three parameters associated with giving-up time and non-directional sensory search strategies. Resource density is 400; cluster radius is 4. Points represent proportional difference in searching efficiency for a single run relative to the mean searching efficiency for the optimal parameter combination.  \(\mu_\text{ext}\) is the extensive movement parameter, \(\mu_\text{int}\) is the intensive movement parameter, GUT is the giving-up time, and SFT is the sensory field threshold. Parameter values are normalized for comparison. Lines are smoothing splines. For both non-directional sensory and GUT search, efficiency generally declines as \(\mu_\text{ext}\) increases, and generally increases as \(\mu_\text{int}\) increases. See A.3 for details.}
\label{fig:splineex}
\end{figure}

For both composite search classes, searching efficiency was most sensitive to movement behavior in extensive mode, \(\mu_\text{ext}\)  (Fig. \ref{fig:splineex}). The difference in searching efficiency between the optimal \(\mu_\text{ext}\) and the worst \(\mu_\text{ext}\) was up to 70\%. In contrast, the difference in searching efficiency between the optimal \(\mu_\text{int}\) and the worst \(\mu_\text{int}\) was no more than 45\%.

Setting the threshold parameter (the time threshold for GUT foragers, the sensory field threshold for non-directional sensory foragers) below the optimal value caused greater decreases in efficiency than when these parameters were set above the optimal value. When the time threshold is set too low, the GUT forager spends too much time in extensive mode; in the extreme, setting the time threshold to zero leads to a reduction in efficiency of nearly 40\%. When the sensory field threshold is set too low, the non-directional sensory forager spends too much time in intensive search; in the extreme, setting this threshold to zero leads to a reduction in efficiency of over 60\% (Figure \ref{fig:splineex}).

Biologically, this means that GUT foragers should error on the side of too much intensive search (with too high a GUT threshold) more frequently than on the side of too little intensive search (with too low a GUT threshold). Conversely, sensory foragers should error on the side of too little intensive search (with too high a sensory field threshold) more frequently than on the side of too much intensive search (with too low a sensory field threshold).

\subsection{Robustness}

\begin{figure}
  \centering
  \includegraphics[width=5 in]{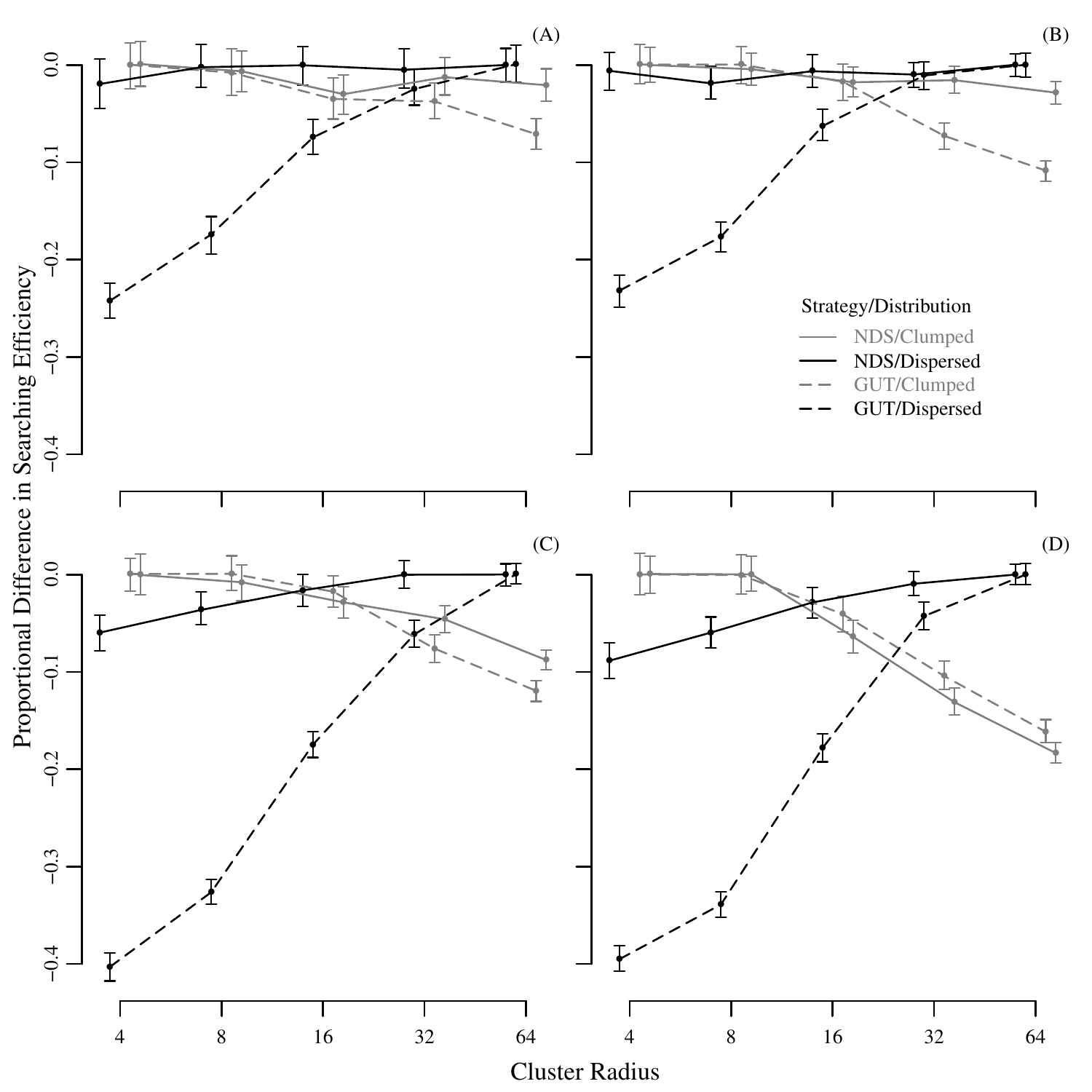}
  \caption{Robustness of non-directional sensory (NDS) and giving-up time (GUT) search strategies across 5 levels of resource aggregation (measured by cluster radii) and 4 levels of resource density: (A) 100, (B) 400, (C) 700, (D) 1000. Robustness measures how well a searcher optimized for a landscape of type X does in a landscape of type Y, relative to a searcher optimized for type Y. A NDS searcher optimized for dispersed landscapes (black solid lines, large cluster radii) performs sub-optimally in clumped landscapes (small cluster radii). The corresponding situation for GUT search is much more dramatic; a GUT searcher optimized for dispersed landscapes (black dashed lines) does very poorly when placed in clumped landscapes. See A.4 for details.}
\label{fig:robustfig}
\end{figure}

Our robustness analysis (explained in detail in \ref{sec:robustness}) allowed us to determine how a forager optimized for a particular level of resource aggregation would fare in landscapes with different levels of resource aggregation. The non-directional sensory strategy was more robust to changes in resource aggregation than the GUT strategy, particularly for foragers that were optimized for dispersed resources (black lines in Fig. \ref{fig:robustfig}). GUT foragers optimized for a high degree of resource aggregation were relatively robust to decreasing degrees of resource aggregation (grey dashed lines in Fig. \ref{fig:robustfig}), but GUT foragers optimized for landscapes with dispersed resources had drastically reduced searching efficiency in landscapes with more aggregated resources (black dashed lines in Fig. \ref{fig:robustfig}). This contrasts with the optimal non-directional sensory foragers, which had much smaller decreases in efficiency (solid lines in Fig. \ref{fig:robustfig}).

Overall, the biological implications of the robustness analysis are clear: non-directional sensory foragers fare better than GUT foragers when placed in environments they are not optimized for. This implies that non-directional sensory foragers will suffer a smaller reduction in efficiency than their GUT counterparts if environments change drastically over time.

\section{Discussion}
\label{sec:4}

Composite search strategies, which consist of extensive and intensive search modes, help foragers focus search effort on resource rich regions and devote less effort to resource poor regions. The central objective of this study was to compare the efficiency of two possible criteria for switching search modes: giving-up time (GUT) and non-directional sensory cues. Our simulations revealed that non-directional sensory foragers outperformed GUT foragers across a full spectrum of resource distributions, ranging from highly aggregated to highly dispersed. In addition, non-directional sensory foragers were more robust to changes in resource distribution, implying that they would be better able to cope with environmental change. These results indicate that it is better to inform search behavior with a non-directional sensory cue than with resource encounters and elapsed time. Together with empirical evidence indicating that sensory cues are more important than recent resource encounters in determining foraging mode \citep{Persons:1997td}, our simulations suggest that the default assumption that GUT governs composite random search should be reexamined. If a researcher in the field observes foragers engaged in composite random search, the default assumption should be the foragers are using the most efficient mode switching strategy, which our simulations show is based on non-directional sensory cues. The alternative hypothesis, that foragers use GUT mediated search, should be entertained only after the researcher rules out all possible sensory cues.

Our simulations also agree with the results in \citet{Benhamou:2007vka}, which showed that composite search strategies outperform non-composite searches in patchy environments. Our simulations show how this performance advantage varies across a spectrum of levels of resource aggregation (Fig. \ref{fig:normeff}). At the lowest levels of resource aggregation, the performance advantage vanishes for GUT composite foragers, but not for non-directional sensory composite foragers.

To our knowledge, GUT is the only mode-switching mechanism previously used to model composite strategies in the general random search context \citep[e.g.,][]{James:2011kw,Reynolds:2009es,Plank:2008gr}. Our model with mode-switching based on non-directional sensory cues is novel. Our results agree with the system-specific model of Hill et al. \citeyearpar{Hill:2003tf}, who simulated juvenile flatfish foraging for bivalves. In their study, simulated flatfish movement was determined by sampling empirically observed movement distributions. The authors studied both a giving-up time composite search strategy and a ``local density" strategy, in which search mode was based on the number of prey items within a fixed radius of the fish's position. Hill et. al found that the local density strategy outperformed the GUT strategy, which matches our finding that the non-directional sensory strategy outperforms the GUT strategy. Our work, which represents very general search behavior, and that of Hill et al. \citeyearpar{Hill:2003tf}, which focused on a specific system, provide complementary evidence for the potential importance of composite foraging strategies that are not based on time.

In an attempt to keep our model as general as possible, we did not include several important ecological factors. First, we did not consider the costs involved in the evolution or development of the cognitive and sensory abilities foragers would need to detect non-directional cues versus the cost to keep track of time. Second, we did not model the process of selection and evolution; an approach similar to that implemented by \citet{Preston:2010eo} could be used to investigate the evolution of these strategies. Third, we only considered non-directional sensory fields that were created by symmetric Gaussian sources. This provides an accurate representation of sensory fields when there are no prevailing winds, as in the Tsetse fly system described by \citet{williams1994ms}. In many systems, however, the sensory field will be altered by wind, currents, and other environmental factors \citep{Reynolds:2012jz}. In some situations, it could be expected that turbulence would blur what otherwise might be used as gradient-following cues, resulting in non-directional sensory cues. In other cases, prevailing winds or straight-line winds might create a situation in which special movement patterns would be optimal \citep{nevitt2008pn}. Future work will be required to fully describe these situations and conditions, but the symmetric Gaussian case studied here provides a useful starting point. Fourth, our simulation was done in two dimensions; for many species, especially marine organisms, a three-dimensional model would be more appropriate. Finally, we did not take into account factors like interspecific competition or predation risk \citep{Brown:2004ko, Reynolds:2010jn}. Information sharing among foragers in a group is another complicating factor.  \citet{Codling:2014ir} created an individual-based model of how leaderless social groups navigate toward a target. Future research should build upon this, and examine how information sharing applies to the context of foraging.

Our models use L\'evy walks to describe movement types that lie on a spectrum between the extremes of Brownian and ballistic motion. Recently, \citet{Pyke:2014gx} argued that traditional L\'evy walk models are biologically unrealistic. He advocated the use of individual based models that can incorporate phenomena like area-restricted search, patchy resource distributions, and forager memory and perceptual abilities. Our modeling framework shows that these features are compatible with L\'evy walks. L\'evy walks provide a simple, general family of non-oriented superdiffusive random walks (for details see, \ref{sec:levyjust}); their usefulness as a descriptive modeling tool does not rest on the validity of the L\'evy foraging hypothesis \citep[as described in][]{Bartumeus:2007tv}. An analogy can be made with Brownian motion, which is frequently used as a descriptive model of random movement, without invoking a Brownian foraging hypothesis. With knowledge of the traits of a particular species and its environment, a highly detailed, specific model can be created; our models, however, aim to explore general principles about composite foraging strategies. We believe L\'evy walks offer an appropriate trade-off between generality and realism.

The modeling framework outlined in this study has the potential to help bridge the gap between two traditionally disparate fields of study: random search theory and classic patch use theory. The former focuses on animal movement patterns, the latter on patch use decisions \citep{Bartumeus:2009fv}. Recent work \citep{Bartumeus:2013vm} has sought to establish a Òstochastic optimal foraging theoryÓ to unify these approaches; our model could contribute to that effort. One of the foundational results of classic foraging theory is Charnov's Marginal Value Theorem (MVT), which dictates that an optimal forager should deplete patches so that the intake rate in each patch is equal to the expected intake rate averaged over the rest of the environment \citep{Charnov:1976wz}. The predictions of the MVT provide a useful benchmark to measure real-world foragers against. Unfortunately, the MVT is not easily translated to the realm of random search theory, where resources have arbitrary spatial distributions (hence patches are not well-defined) and resource encounters are typically discrete events (hence instantaneous intake rate is not well-defined).

\citet{Plank:2008gr} proposed an analogue between patch-use models and composite random search models: within patch harvesting corresponds to intensive search, while between-patch travel corresponds to extensive search. They further suggested that optimal GUT composite searchers represent the random search version of MVT optimal foragers. There are important differences between the optimal behavior predicted by these two models, though. MVT optimal foragers make decisions based on the current local and global resource levels. They are omniscient, and hence have no need to use past experience or memory. This contrasts with GUT optimal foragers, whose behavior is highly dependent on stochastic resource encounters. The non-directional sensory optimal foragers introduced in this paper provide a better analogue to MVT optimal foragers. Like MVT optimal foragers, non-directional sensory optimal foragers make instantaneous assessments of local and global resource conditions to determine when to switch behavioral modes. Just as MVT optimal foragers provide a useful null-model for foraging on landscapes with resource patches, non-directional sensory optimal foragers provide a useful null-model for foraging on landscapes with resources distributed as arbitrary point patterns. Similar to MVT foragers, the non-directional sensory optimal foragers can be used as a benchmark, even when the mechanisms foragers use to locate resources are unknown. The non-directional sensory forager model predicts areas that warrant intensive search; by overlaying this with observed animal movement trajectories, one can determine how close those animals come to optimal behavior.

In summary, our results challenge the assumption that GUT is the key criterion composite foragers use to switch between intensive and extensive search modes. Non-directional sensory cues and GUT are both potential mechanisms for mediating kinesis, but our simulations show that strategies based on non-directional sensory cues are more efficient than strategies based on GUT. This suggests that foragers should rely more heavily on sensory cues than elapsed time for determining search mode switches.

\section{Acknowledgements}

This work was supported in part by NSF DEB 0953766 to CEB. The supercomputing resources provided by the Holland Computing Center at the University of Nebraska-Lincoln greatly facilitated the collection of simulation data. We are grateful to four anonymous reviewers for their helpful comments on this manuscript.

\appendix

\renewcommand*{\thesection}{\Alph{section}}

\section{Appendix}

\subsection{Diffusive and superdiffusive random walks}
\label{sec:levyjust}

A frequently invoked reason for using L\'evy walks to model animal movement is that they are``superdiffusive" \citep{Viswanathan:2008in}. In this section, we clarify how this term is defined, and address the implications for situations where walks are truncated.

Mean-square displacement (MSD) provides a useful way to classify stochastic movement. If a particle's position is given by the stochastic process $x\left(t\right)$, then we define $\text{MSD}=\left\langle x^{2}\right\rangle $.
For Brownian motion, $\left\langle x^{2}\right\rangle \sim t$. This linear scaling of MSD with time is referred to as normal diffusion. If $\left\langle x^{2}\right\rangle \sim t^{\alpha},\:\alpha>1$, then we say that the particle is ``superdiffusive".

For random walks with step-lengths drawn from a Pareto distribution with $\mu>3$, the Central Limit Theorem guarantees that $\left\langle x^{2}\right\rangle  \sim t$. Hence these random walks are normal diffusions, and behave like Brownian motion at sufficiently large time scales. For $\mu=3$, the Central Limit Theorem no longer applies, and $\left\langle x^{2}\right\rangle \sim\ln\left(t\right)t$, a marginal case between normal diffusion and superdiffusion. This fact is mentioned in \citet{klafter1996bd} and \citet{Viswanathan:2011vl}, and a proof of the discrete space case can be found in \citet{Zumofen:1993aa}. We were able to translate Zumofen and Klafter's approach into a continuous space context (we were unable to find a proof of this case in the literature). For $\mu<3$, it is well known that the resulting L\'evy walk is superdiffusive \citep{Viswanathan:2011vl}.

These results are valid for theoretical L\'evy walks. In nature and in simulations, there will always be an upper bound to step-lengths. These limitations result in truncated L\'evy walks, which, by the Central Limit Theorem, eventually converge to Brownian motion \citep{Benhamou:2007vka}. There are both numerical and philosophical reasons why this does not prevent L\'evy walks from serving as useful modeling tools. The numerical argument, as established in \citet{Mantegna:1994vc} and reiterated in \citet{Viswanathan:2011vl}, is that the convergence to Brownian motion is very slow, and hence a truncated L\'evy walk retains its superdiffusive quality over relevant time scales. From a philosophical perspective, the theoretical (non-truncated) L\'evy walk represents a process that, together with the environment, generates an observed (or simulated) truncated L\'evy walk. As argued by \citet{Viswanathan:2011vl}, this distinction between process and observation is extremely important.

Another important property of L\'evy walks is that they are scale-free; that is, the probability distribution of step lengths is invariant under scaling transformations \citep{Reynolds:2009ub}. When a forager interacts with its environment, either by truncating step lengths or by switching its search mode, the scale-free property is not retained \citep{Viswanathan:2008in}. In other words, although scale-invariant probability distributions drive a forager's intrinsic movement tendencies, its trajectory depends on its interactions with the environment, and hence is not scale-free. This is realistic, because foraging behavior should reflect the characteristic scales of the surrounding environment.

\subsection{Parameter optimiziation}
\label{sec:optimize}

We used a grid-based search to explore the searching efficiency associated with large regions of the parameter space of our simulation model. A non-composite forager is characterized by a single parameter \(\mu\). We ran non-composite simulations using parameter values \(\mu=1.0, 1.2, 1.4, ..., 3.0\) on each landscape type (specified by initial resource distribution and resource aggregation). For the composite foragers, we examined 4 initial resource densities, 5 cluster radii, 2 search strategies (GUT and non-directional sensory), and 11 values for each of the 3 search parameters (\(\mu_\text{ext}\), \(\mu_\text{int}\), switching threshold). In the first sweep of the parameter space, we conducted 100 runs for each parameter combination for a total of 5,324,000 runs (4 densities * 5 radii * 2 strategies * $11^{3}$ search parameter combinations * 100 runs). Each run of the model consisted of 20,000 discrete time steps. Even the most efficient foragers did not come close to consuming all of the available resources; hence totally depleting the landscape before 20,000 time steps was not an issue. The full grid-based search produced a rough fitness surface based on the searching efficiency of each parameter combination. The fitness surface allowed us to exclude regions of the parameter space that led to poor searching efficiency, thereby focusing our computational resources on increasing replication in regions of the parameter space that were likely to contain the optimal parameter combination. We used an iterative process (described below) to narrow the regions of the parameter space selected for increased replication. The iterative process did not produce a finer-scale resolution of the parameter space but rather increased the replication for subsets of the parameter combinations used in the full grid-based search. Within each landscape type, we used the mean searching efficiency from the full grid-based search to select the top 13 of the 1331 (1\%) possible parameter combinations. For each parameter, we used the range of values found within the top 1\% to reduce the parameter space. For example, suppose the top 1\% parameter combinations included \(\mu_\text{ext}\) values that ranged from 1.0-1.4, \(\mu_\text{int}\) values from 2.6-3.0, and GUT values from 100-200. Then we would have increased replication for the 27 parameter combinations (\(\mu_\text{ext}\), \(\mu_\text{int}\), GUT) that represented parameter values within those ranges: $\mu_\text{ext}=1.0,\,1.2,\,1.4$; $\mu_\text{int}=2.6,\,2.8,\,3.0$; $\mbox{GUT}=100,\,150,\,200$. For some landscape types, this approach did not reduce the parameter space substantially. Thus, we conducted 200 runs for each parameter combination in the reduced parameter space and again calculated the top 1\% of the parameter combinations to further reduce the parameter space. This process was repeated until the optimal parameter combination was comprised of at least 500 runs because preliminary exploration of the model indicated that 500 runs produced good estimates of mean searching efficiency.

\subsection{Sensitivity}
\label{sec:sensitivity}

We examined the sensitivity of searching efficiency to each search parameter by varying one search parameter while holding the other two parameters at their optimal values. \(\mu_\text{ext}\) and \(\mu_\text{int}\) ranged from 1 to 3, GUT ranged from 0 to 500, and the sensory field threshold ranged from 0 to 0.256 (Table \ref{tab:parmvals}). The \(\mu\) parameters have a naturally bounded range, but the threshold parameters have arbitrary upper bounds, which were selected based on preliminary explorations of parameter space. We normalized the parameter values to fall between 0 and 1 to facilitate comparisons across the different ranges of the parameters. We calculated the proportional difference in searching efficiency as \(D_\text{S} = (y-\bar{y}_\text{o})/\bar{y}_\text{o}\), where \(y\) was the searching efficiency for a single run and \(\bar{y}_\text{o}\) was the mean searching efficiency for the optimal parameter combination. We fitted smoothing splines to the relationship between \(D_\text{S}\) and the normalized value of each parameter for each landscape type. The shape of the smoothing splines provided an indication of the sensitivity of searching efficiency to changes in each parameter. In two cases (see Table \ref{tab:optparms}), the optimal \(\mu_\text{ext}\)  and \(\mu_\text{int}\) were the same, which made the best giving-up time parameter irrelevant. Thus, those landscape types were excluded from the sensitivity analysis.

\subsection{Robustness}
\label{sec:robustness}
To assess the robustness of the optimal strategies to changes in resource aggregation, we examined how a search strategy that maximized the searching efficiency for one landscape type performed in landscape types with different degrees of resource aggregation. Specifically, we calculated robustness as \(D_{R}=(\bar{y}_{i,j}-\bar{y}_{i,i})/{\bar{y}_{i,i}}\), where \(\bar{y}_{i,j}\) was the mean searching efficiency in landscapes of type \(i\) for a forager that was optimized for a landscape of type \(j\).  In this formula, landscape types are indexed by cluster radius. We examined how foragers optimized for very clumped and very disperse landscapes (\(j=4\) and \(j=64\), respectively) performed on a full range of landscape types (\(i=4, 8, 16, 32, 64\)). This analysis was done on four different levels of resource density (100, 400, 700, 1000). Then we resampled the data with replacement (i.e., bootstrap method) 500 times for each landscape type and calculated the mean and 2.5\% and 97.5\% quantiles of the distribution of robustness values. 

\subsection{Boundary conditions}
\label{sec:boundary}

Landscape boundary conditions play an important role in individual-based models \citep{Berec:2002tx}. Most simulations use one of three types of boundary conditions: reflecting, periodic, or absorbing. Reflecting boundaries are appropriate for modeling animals that live in a restricted environment, like animals on an island \citep{Berec:2002tx}, or animals with territories bounded by scent marks \citep{Giuggioli:2014tc}. Reflecting boundary conditions can also be interpreted as having a new forager enter the landscape at the exact place where the previous forager left it. This biases the initial conditions for the new forager and creates edge effects.

Periodic boundary conditions can be interpreted in three different ways. First, the landscape is literally a torus, which is an unrealistic assumption. Second, the landscape is infinite, but repeating; this is problematic when resource consumption is destructive, and a forager's actions at one point on a landscape affect an infinite number of other points. Third, a new forager enters the landscape at a point determined by where the original forager left it; like with reflecting boundary conditions, this has the potential to create edge effects. Our modeling framework presents a few additional problems associated with periodic boundary conditions. The resource distributions and the sensory field are generated under the assumptions that the topology of the landscape is a plane; periodic boundary conditions would mean that resources on opposite ends of the landscape are close to each other, leading to logical inconsistencies.

Alternatively, foragers that reach the boundary of the landscape could be reintroduced at a randomly selected point on the boundary. This poses a problem, because a random walker placed on a boundary is highly likely to cross the boundary within a very short time period (often immediately). This can lead to a long repeated pattern of foragers being placed on the boundary, crossing it almost immediately, being replaced on the boundary, leaving again, etc. Ideally, one would like to de-emphasize the role of potential boundary artifacts, and minimize the time a forager spends near boundaries. The boundary reintroduction method does not meet that goal.

In our model, we implemented a modified version of absorbing boundary conditions. The major challenge with absorbing boundary conditions is that a forager could leave the landscape by chance almost immediately after entering it. The performance of such a forager would not provide much information about the efficiency of the strategy it employed. Therefore, we chose to force each forager to spend 20,000 discrete time steps foraging on the landscape. If the forager was absorbed by a boundary, it was randomly dropped back into the landscape to resume foraging. This can be interpreted as a forager leaving the landscape, then returning later to resume foraging. We chose 20,000 time steps, because this was a sufficient time for foragers to appreciably deplete landscapes. Finally, we included a small resource-free buffer zone at the edge of the landscape. The entire landscape (including buffers) was a square 111 units long and 111 units wide, but only the 101 unit long, 101 unit wide square in the center contained resources. Resource-free buffer zones occupied 5 unit thick strips at the top, bottom, left, and right edges of the central landscape. This ensured that all resources could be approached from every direction, and that no resources were protected by edge effects.

\bibliographystyle{model2-names}
\bibliography{CSReferences.bib}

%% not want to use model2-names.bst.



%% \bibitem must have one of the following forms:
%%   \bibitem[Jones et al.(1990)]{key}...
%%   \bibitem[Jones et al.(1990)Jones, Baker, and Williams]{key}...
%%   \bibitem[Jones et al., 1990]{key}...
%%   \bibitem[\protect\citeauthoryear{Jones, Baker, and Williams}{Jones
%%       et al.}{1990}]{key}...
%%   \bibitem[\protect\citeauthoryear{Jones et al.}{1990}]{key}...
%%   \bibitem[\protect\astroncite{Jones et al.}{1990}]{key}...
%%   \bibitem[\protect\citename{Jones et al., }1990]{key}...
%%   \harvarditem[Jones et al.]{Jones, Baker, and Williams}{1990}{key}...
%%


% \end{thebibliography}

\newpage

\renewcommand{\thefigure}{\arabic{figure}}
\renewcommand{\thetable}{\arabic{table}}

\end{document}